\documentclass{ws-ijbc1}

\usepackage{xcolor}

\usepackage[verbose]{hyperref}
\hypersetup{colorlinks=false,allbordercolors=blue,pdfborderstyle={/S/U/W 1}}

\begin{document}

\catchline{}{}{}{}{} 

\markboth{Nuumata, Saito}{Multiobjective Optimization}

\title{A Fundamental Study for Multiobjective Optimization Problems in Nonlinear Dynamical Systems}

\author{Ryunosuke Numata and Toshimichi Saito}
\address{Department of Electorical and Electronc Engineering, HOSEI University\\
Koganei, Tokyo, 184-8584, Japan
\\tsaito@hosei.ac.jp}

\maketitle

\begin{history}
\end{history}

\begin{abstract}
Multiobjective optimization problems are important in analysis and application of nonlinear dynamical systems.  
As a first step, this paper studies a  biobjective optimization problem in a simple nonlinear switched dynamical system: a piecewise linear system based on a boost converter with photovoltaic input. 
The piecewise linearity enables us to analyze the nonlinear dynamics exactly. 
In the biobjective optimization problem, the first objective evaluates stability of circuit operation and the second objective evaluates average input power.  
A main task is analysis of a trade-off between the two objectives. 
Using the piecewise exact solutions, the two objectives are formulated theoretically. 
Using the theoretical formulae, the existence of a trade-off between the two objectives is clarified exactly. 
Relationship between the trade-off and parameters is also considered. 
The results provide fundamental information to analyze multiobjective optimization problems in various nonlinear systems and to realize their engineering applications. 
\end{abstract}


\vspace*{-0.5in}
\section{Introduction}
\label{intro}
Multiobjective optimization problems (MOPs) require simultaneous optimization of multiple objectives, e.g. stability and efficiency. 
In the MOP, we often encounter a trade-off: as one objective improved, another objective deteriorates. 
In order to analyze the trade-offs, efficient evolutionary algorithms have been presented and the algorithm efficiency has been evaluated in benchmarks such as the multiobjective 0-1 knapsack problems \cite{mop1} \cite{mop2} \cite{mop3}.  
The MOPs have been studied in various systems including deep neural networks and power supply in electric vehicles \cite{pf1} \cite{pf2} \cite{pf3}. 
However, the MOPs in nonlinear dynamical systems have not been studied sufficiently. 
Especially, exact analysis of the MOPs is not sufficient. 
The reasons include difficulty in calculation of objective functions based on nonlinear dynamics. 

As a first step to consider the MOPs in nonlinear dynamical systems, this paper studies a biobjective optimization problem (BOP) in a simple switched dynamical system (SDS, \cite{sds1} \cite{sds2}). 
The SDSs are continuous-time nonlinear dynamical systems with continuous and discrete state components.
If the state variable fulfills a certain condition, a switching event occurs which controls the further dynamics. 
The SDSs have been studied extensively in analysis of nonlinear phenomena and in modeling of nonlinear systems such as recurrent neural networks and switching power converters \cite{sds3} \cite{sds4}\cite{sds5} \cite{sds6}. 

\begin{figure}[b]
\centering
\includegraphics[width=0.6\columnwidth]{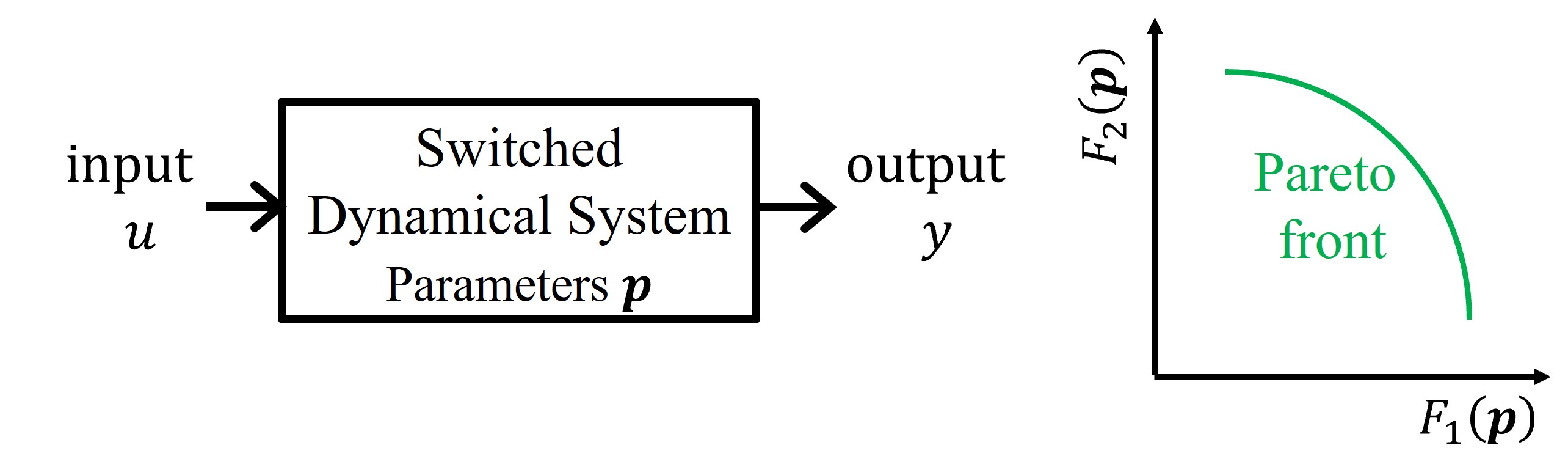}
\caption{Biobjective optimization problems in nonlinear  switched dynamical systems.}
\label{fg1} 
\end{figure}

\begin{figure}[b]
\centering
\includegraphics[width=0.6\columnwidth]{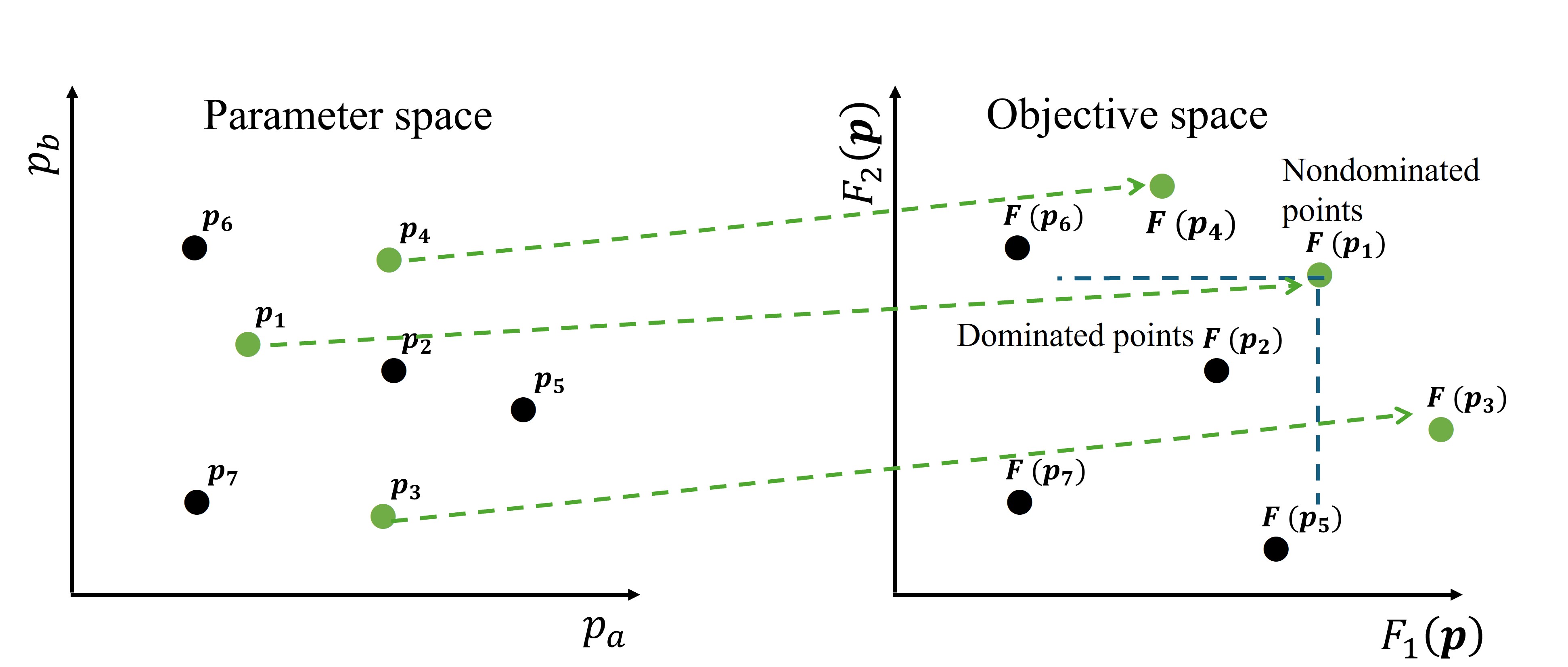}
\caption{Domination concept and  Pareto front 
$(\bm{F}(\bm{p_1}), \bm{F}(\bm{p_3}), \bm{F}(\bm{p_4}))$.}
\label{fg2} 
\end{figure}

Fig. \ref{fg1} illustrates a conceptual diagram of a BOP in a SDS having parameters $\bm{p}$. 
The BOP requires simultaneous optimization (maximization) of the two objective functions of $\bm{p}$: 
\begin{center}
Maximize $\bm{F} \equiv (F_1(\bm{p}), F_2(\bm{p})) \in S_o$, 
subject to $\bm{p} \in S_p$
\end{center}
where $S_o$ denotes the objective space and $S_p$ denotes the parameter space. 
In the BOP, if a trade-off exists between the two objectives, the trade-off is characterized by Pareto front based on domination concept as illustrated in Fig. \ref{fg2}.
That is, a parameter value $\bm{p_1} \in S_p$ is said to dominate another value $\bm{p_2} \in S_p$, if the following is satisfied: 
\begin{equation}
\begin{array}{lll}
(F_1(\bm{p_1}) > F_1(\bm{p_2})  & \mbox{ and }  & F_2(\bm{p_1}) > F_2(\bm{p_2})), \\
(F_1(\bm{p_1}) > F_1(\bm{p_2})  & \mbox{ and }  & F_2(\bm{p_1}) = F_2(\bm{p_2})), \mbox{  or }\\
(F_1(\bm{p_1}) = F_1(\bm{p_2}) & \mbox{ and }  & F_2(\bm{p_1}) > F_2(\bm{p_2}))
\end{array}
\label{ndom}
\end{equation}

\noindent
A parameter value $\bm{p_n}$ is referred to as an optimal solution if $\bm{p_n}$ is not dominated by any other parameter values in $S_p$. 
The set of all the optimal solutions is called Pareto set while its image in the objective space is called Pareto front \cite{mop1}. 
The Pareto front characterizes the best trade-off. 
The BOP is important in fundamental study such as analysis of a SDS performance (e.g., stability and efficiency) for the parameters. 
It is also important in applications such as design of a SDS with optimal performance. 
It should be noted that the BOPs of the SDSs require analysis in both the parameter space and the objective space. 
On the other hand, stability/bifurcation of  the SDSs is analyzed in the parameter space.  

As a concrete example of the SDSs, this paper considers a simple piecewise linear system based on a boost converter with photovoltaic (PV) input. 
The boost converters have been studied extensively in renewable energy supply systems in carbon neutral technology \cite{boost1} \cite{boost2} \cite{boost3}. 
The piecewise linearity enables us to analyze nonlinear dynamics exactly \cite{jetcas} \cite{fujikawa} \cite{apccas}. 

In the piecewise linear system, we define the BOP where the first objective evaluates stability of circuit operation and the second objective evaluates average power from the PV input. 
We give two main results. 
First, using the piecewise exact solutions, the two objectives are formulated theoretically. 
Using the formulae, the two objectives are calculated exactly on a two-dimensional parameter subspace. 
Relationship between the parameter subspace and objective space is considered. 
Second, applying the brute force attack to the theoretical formulae on the parameter subspace, we obtain the Pareto front: the existence of a trade-off between the two objectives is guaranteed. 
These results are exact without approximation. 

These results are fundamental to consider MOPs in various nonlinear dynamical systems and their applications.  
It should be noted that this is the first paper of exact analysis of BOPs in nonlinear dynamical systems. 
Preliminary results along these lines can be found in \cite{apccas}.

\section{The Simple Switched Dynamical System}
\label{pwl}
\begin{figure}[b]
\centering
\includegraphics[width=0.6\columnwidth]{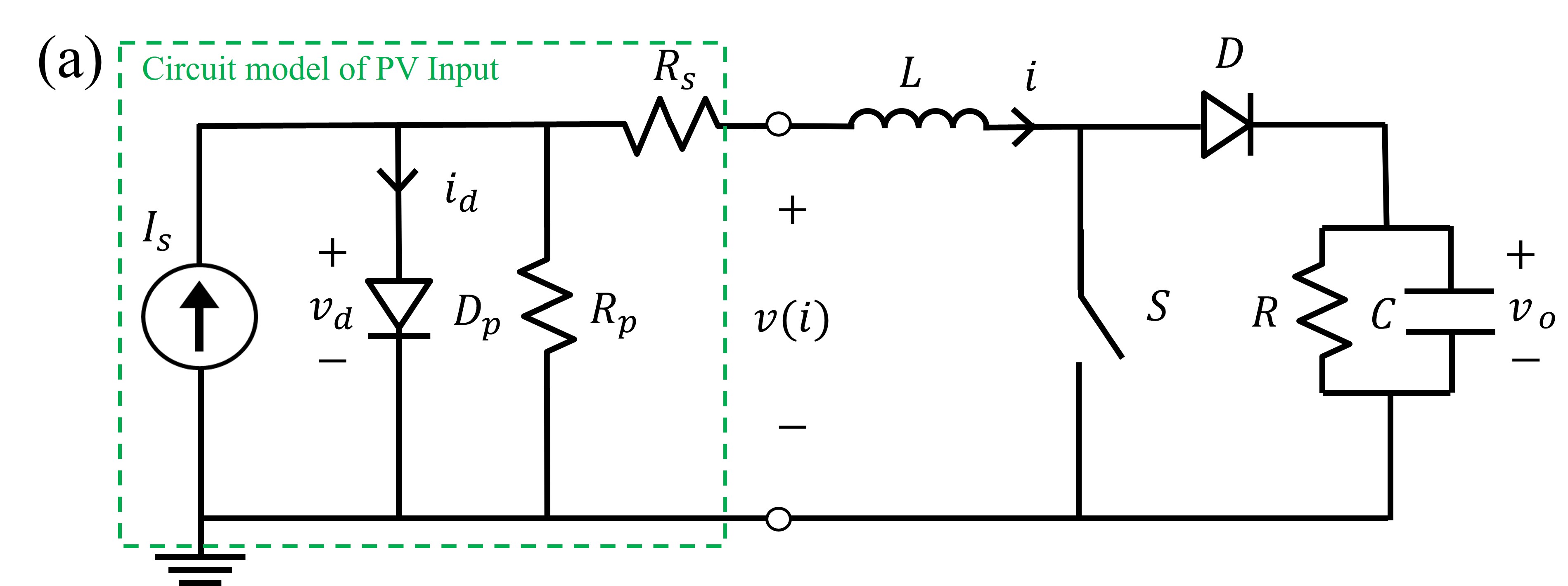}
\includegraphics[width=0.6\columnwidth]{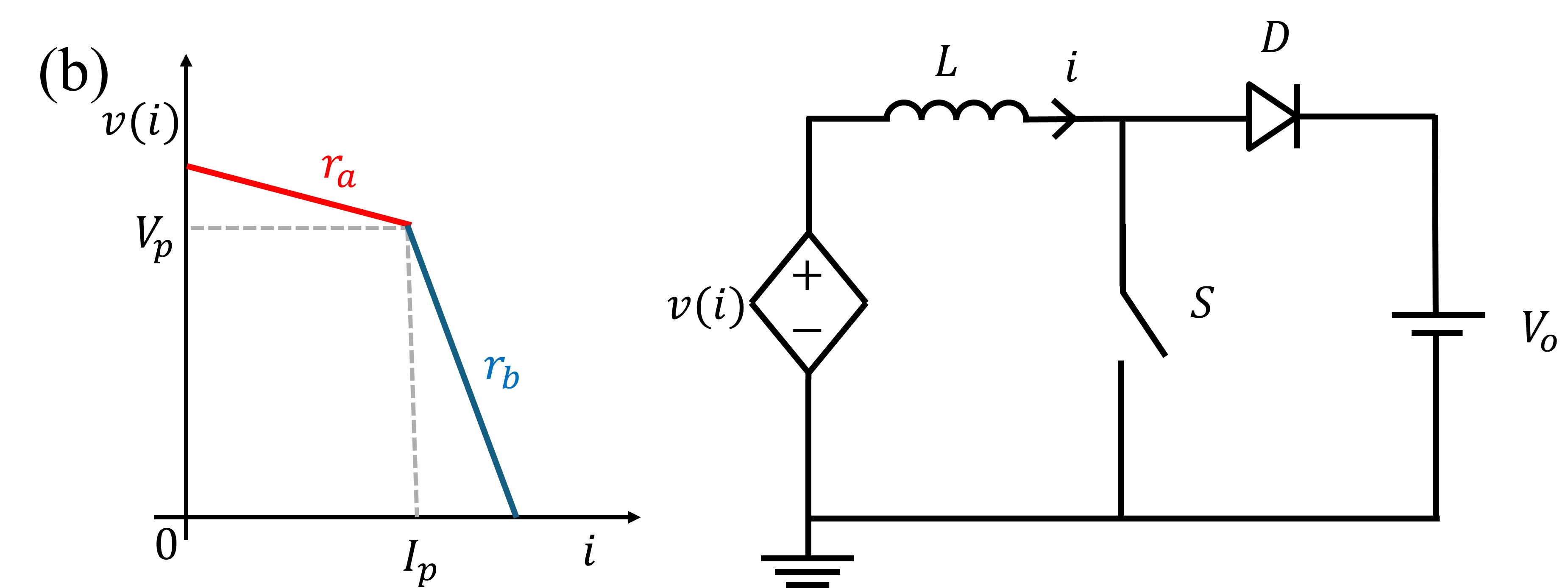}
\includegraphics[width=0.6\columnwidth]{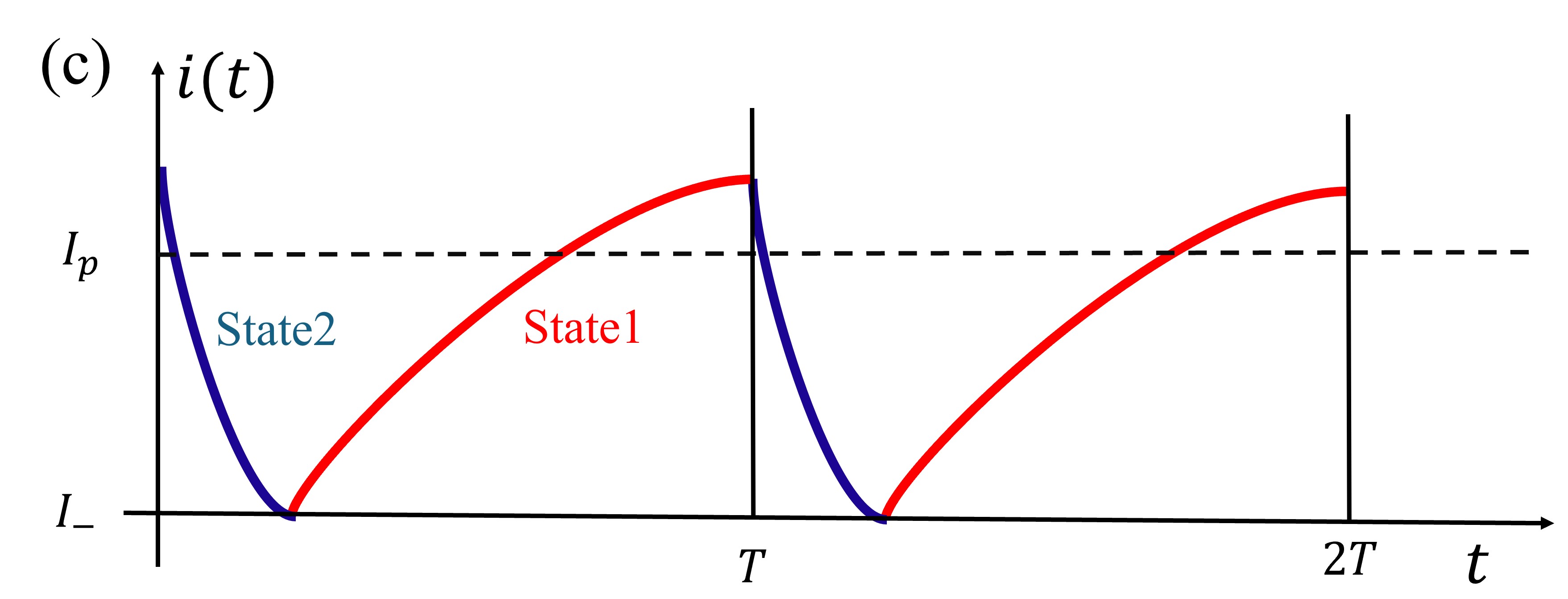}
\caption{Boost converter with PV input. (a) Circuit model. (b) Piecewise linear model. (c) Switching rule}
\label{fg3} 
\end{figure}

This section introduces the simple piecewise linear SDS.  
Fig. \ref{fg3} shows a circuit model of boost converter with the PV input.  
The PV input is expressed by an equivalent circuit \cite{pv1} \cite{pv2}:  
the current source $I_s$ is photo-generated current, 
the diode $D_p$ models the p-n junction of the solar cell, 
$R_s$ is series resistance, and $R_p$ is shunt resistance.  
Referring to \cite{apccas}, the diode is simplified into the  piecewise linear element 
\begin{equation}
    \begin{array}{l}
    i_d(v_d) = \left\{
        \begin{array}{ll}
        g_d(v_d - V_T) & \mbox{for $v_d \geq V_T$}\\
        0 & \mbox{for $v_d < V_T$}
        \end{array}\right.
    \end{array}
\end{equation}
Applying KCL/KVL, we obtain the piecewise linear characteristic of the PV input: 
\begin{equation}
    \label{CCVS}
    \begin{array}{l}
        \displaystyle v(i) =
        \left\{\begin{array}{ll}
            -r_a(i-I_p)+V_p      & \mbox{for $0<i < I_p$}\\
            -r_b(i-I_p)+V_p      & \mbox{for $I_p \leq i$}
        \end{array}\right. \\
    \end{array}
\end{equation}
where $r_a \equiv R_s + R_p/(1+g_dR_p)$, $r_b \equiv R_s + R_p$, 
$I_p \equiv I_s - V_T/R_p$, and $V_p \equiv (1+R_s/R_p)V_T - R_sI_s$. 
This PV input is regarded as the current-controlled voltage source (CCVS).  

We define the switching rule. 
First, the switch $S$ and diode $D$ can be either
\[
\begin{array}{l}
\mbox{State 1 ($S$ conducting,  $D$ blocking,  and $I_- \le i$)} \mbox{ or}\\
\mbox{State 2 ($S$ blocking,  $D$ conducting,  and $I_- \le i$)}
\end{array}
\]
where $I_-$ is the lower threshold of inductor/input current $i$. 
Second, let $T$ be the period of the clock signal. 
As illustrated in Fig. \ref{fg3} (c),  in State 1, the inductor current $i$ rises. 
The State 1 is switched into State 2 at $t=nT$. 
In State 2, $i$ decreases.
The State 2 is switched into State 1 when  $i$ reaches $I_-$. 

\begin{figure}[b]
\centering
\includegraphics[width=0.6\columnwidth]{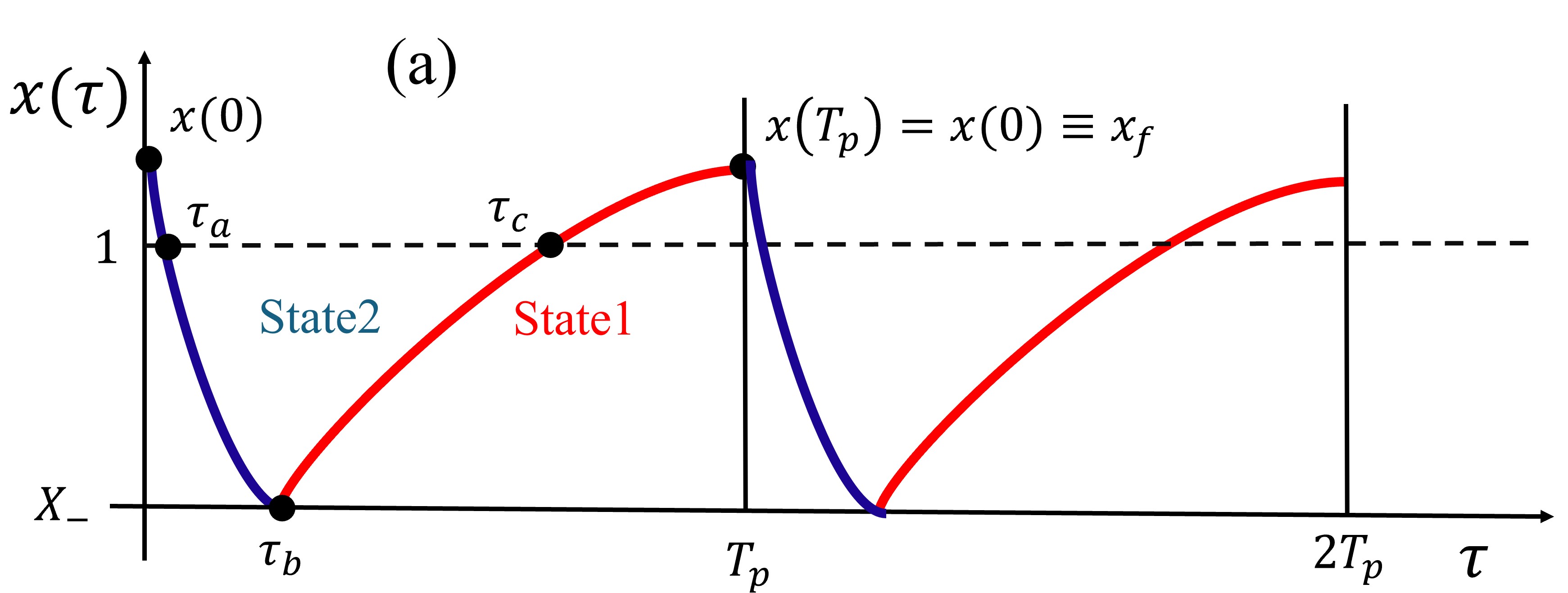}
\includegraphics[width=0.6\columnwidth]{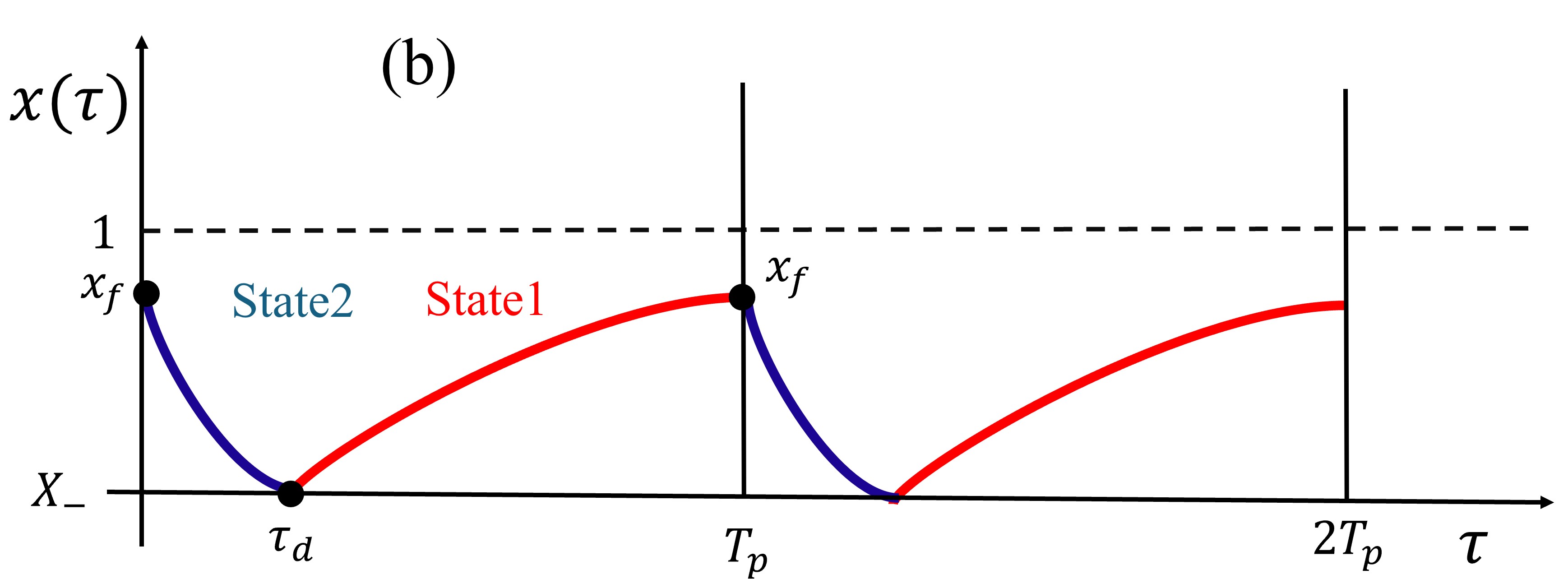}
\caption{Periodic orbit with period $T_p$. (a) Type 1. (b) Tyoe 2.}
\label{fg4} 
\end{figure}

We assume that time constant of the load is large enough, $RC \gg T$. 
Referring to \cite{jetcas}, the load is simplified into the constant voltage source $V_o$. 
The circuit dynamics is described by 
\begin{equation}
    \label{ODE}
    \begin{array}{l}
       \displaystyle L\frac{di}{dt} =
        \left\{\begin{array}{ll}
            v(i)             & \mbox{for State1}\\
            -(V_o-v(i))      & \mbox{for State2}
        \end{array}\right. 
    \end{array}
\end{equation}
The instantaneous input power is given by $P(t) = iv(i)$. 
Using dimensionless variables and parameters:   
\begin{equation}
    \label{parameter}
\begin{array}{c}
\displaystyle \tau = \frac{V_p}{LI_p}t, \ x = \frac{i}{I_p}, \ 
y = \frac{v(i)}{V_p}\\[5mm]
\displaystyle T_p = \frac{V_p}{LI_p}T, \  \alpha = \frac{I_p}{V_p}r_a, \ \beta = \frac{I_p}{V_p}r_b, \  X_- = \frac{I_-}{I_p}, \  q = \frac{V_o}{V_p},
\end{array}
\end{equation}
Eq. (\ref{ODE}) is transformed into the dimensionless piecewise linear SDS 
\begin{equation}
    \label{DODE}
    \begin{array}{l}
       \displaystyle \frac{dx}{d\tau} =
        \left\{\begin{array}{ll}
            y(x)      & \mbox{for State1}\\
            y(x)-q      & \mbox{for State2}
        \end{array}\right. \\[5mm]
\mbox{State 1 $\longrightarrow$ State 2 when $\tau=nT_p$}\\
\mbox{State 2 $\longrightarrow$ State 1 when $x=X_-$}
    \end{array}
\end{equation}
where $y(x)$ is the dimensionless PV input 
\begin{equation}
y(x) =\left\{\begin{array}{ll}
            -\alpha(x-1)+1      & \mbox{for $0<x<1$}\\
            -\beta(x-1)+1      & \mbox{for $1 \leq x$}
        \end{array}\right.
\label{io}
\end{equation}
The dimensionless instantaneous input power is given by
$p(\tau) = x(\tau)y(\tau)$. 
This dimensionless piecewise linear SDS is the concrete subject of the BOP in Section \ref{bop}. 
This SDS is a rough model and cannot describe practical boost converters precisely, however, the SDS enables as to calculate the objectives and Pareto front exactly as shown afterword. 
The exact analysis is important in fundamental studies of nonlinear dynamics and
provides fundamental information for considering practical systems. 
It should be noted that the dimensionless five parameters ($T_p$, $\alpha$, $\beta$, $X_-$, $q$) can cope with various specifications based on practical circuit parameter values. 
Depending on the dimensionless parameters, the SDS can exhibit various nonlinear phenomena: periodic orbits, chaos and bifurcation \cite{fujikawa}.  
For convenience in analysis of the BOPs. we focus on the periodic orbits with period $T_p$, $x(\tau + T_p) = x(\tau)$, as shown in Fig. \ref{fg4}.

\section{Periodic Orbits and Theoretical Formulae}

This section gives theoretical formulae of the periodic orbits for the BOP. 
The periodic orbits are classified into two types. 
In Type 1, as shown in Fig. \ref{fg4} (a), the orbit $x(\tau)$ starts from an initial point $x(0) > 1$ at $\tau=0$ in State 2. 
$x$ decreases and reaches the break point $x=1$ at time $\tau_a$. 
$x$ decreases and hits the lower threshold $X_-$ at time $\tau_b$ at which the State 2 is switched into State 1. 
$x$ increases and reaches the break point $x=1$ 
at time $\tau_c$.  
$x$ increases and reaches the point $x(T_p)$ at time $T_p$ at which State 1 is switched into State 2.  
In the periodic orbit, $x(T_p)=x(0) \equiv x_f$ is satisfied. 
Using the exact piecewise solutions, periodic orbit is described exactly: 
\begin{equation}
\begin{array}{ll}
\left\{\begin{array}{l}
x(\tau) = (x(0) - B_2)e^{-\beta \tau} + B_2\\[2mm]
\displaystyle x(\tau_a) = 1: \ \beta \tau_a = \ln\frac{x(0)-B_2}{1-B_2} \end{array}\right. 
 & \mbox{for }  0 < \tau  \leq \tau_a\\[7mm]
\left\{\begin{array}{l}
x(\tau) = (1 - A_2)e^{-\alpha(\tau - \tau_a)} + A_2\\[2mm]
\displaystyle x(\tau_b) = X_-: \ \alpha \tau_b = \ln\frac{1-A_2}{X_--A_2} + \alpha \tau_a \end{array}\right.
 & \mbox{for }  \tau_a < \tau  \leq \tau_b\\[7mm]
\left\{\begin{array}{l}
x(\tau) = (X_- - A_1)e^{-\alpha(\tau - \tau_b)} + A_1\\[2mm]
\displaystyle x(\tau_c) = 1: \ \alpha \tau_c = \ln\frac{X_--A_1}{1-A_1} + \alpha \tau_b
\end{array}\right. 
 & \mbox{for }  \tau_b < \tau  \leq \tau_c\\[7mm]
\hspace*{4mm} x(\tau) = (1 - B_2)e^{-\beta(\tau - \tau_c)} + B_2 
 & \mbox{for }  \tau_c < \tau  \leq T_p
\end{array}
\label{orbit}
\end{equation}
\[
A_1 \equiv (\alpha+1)/\alpha, 
A_2 \equiv (\alpha+1-q)/\alpha, 
B_1 \equiv (\beta+1)/\beta, 
B_2 \equiv (\beta+1-q)/\beta. 
\]
The initial point $x_f$ of the Type 1 periodic orbit is given by
\begin{equation}
x_f = f_1(x_f), \ 
x(T_p) =  f_1(x(0)) \equiv Df_1(x(0) - B_2) + B_1\\
\label{peo1}
\end{equation}
\begin{equation}
Df_1 \equiv \left(\frac{(X_--A_1)(1-A_2)}{(1-A_1)(X_--A_2)} \right)^{\frac{\beta}{\alpha}}\frac{1-B_1}{1-B_2}e^{-\beta T_p}
\label{df1}
\end{equation}

\noindent
In Type 2, as shown in Fig. \ref{fg4} (b), the orbit $x(t)$ starts from an initial point $x(0) < 1$ in State 2. 
$x$ decreases and hits the lower threshold $X_-$ at time $\tau_d$ at which the State 2 is switched into State 1. 
$x$ increases and reaches the point $x(T_p)$ at time $T_p$ 
at which State 1 is switched into State 2.  
Using the exact piecewise solutions, the Type 2 periodic orbit is described exactly: 
\begin{equation}
\begin{array}{ll}
\left\{\begin{array}{l}
x(\tau) = (x(0) - A_2)e^{-\alpha \tau} + B_2\\[2mm]
\displaystyle x(\tau_d) = X_-: \ \alpha \tau_d = \ln\frac{x(0)-A_2}{X_--A_2}\end{array}\right. 
 & \mbox{for }  0 < \tau  \leq \tau_d\\[7mm]
\hspace*{4mm} x(\tau) = (X_- - A_1)e^{-\alpha(\tau - \tau_d)} + A_1  & \mbox{for }  \tau_d < \tau  \leq T_p
\end{array}
\label{orbit2}
\end{equation}
The initial point $x_f$ of the Type 2 periodic orbit is given by
\begin{equation}
x_f = f_1(x_f), x(T_p) =  f_2(x(0)) \equiv Df_2(x(0) - A_2) + A_1
\label{peo2}
\end{equation}
\begin{equation}
Df_2 \equiv \frac{X_- -A_1}{X_- - A_2}e^{-\alpha T_p}
\label{df2}
\end{equation}
The initial point $x_f$ is the fixed point of function $f_1$ or $f_2$ and is given by
\begin{equation}
x_f = \left\{\begin{array}{ll}
\displaystyle \frac{B_1 - Df_1 B_2}{1-Df_1} & \mbox{ for Type 1}\\[5mm]
\displaystyle \frac{A_1 - Df_2 A_2}{1-Df_2} & \mbox{ for Type 2}
\end{array}\right.
\label{fpoint}
\end{equation}
where we ignore the cases $|Df_1| = 1$ and $|Df_2| = 1$. 
The periodic orbit is stable if the contraction rate satisfies $|Df_1|<1$ or $|Df_2|<1$. 
The theoretical formulae (\ref{df1}) and (\ref{df2}) are used in analysis of the BOP. 

The dimensionless average input power of the periodic orbit ( the average input power) is also described exactly. 
In the Type 1 periodic orbit, we obtain
\begin{equation}
P_{A1} = \frac{1}{T_p}\int_0^{T_c} p(\tau) d\tau = \frac{1}{T_p}\int_0^{T_c} x(\tau)y(\tau) d\tau = \frac{q}{T_p}\left(\frac{1-X_-}{\alpha}+\frac{x_f-1}{\beta}+A_2(\tau_b-\tau_a)+B_2\tau_a\right)
\label{pa1}
\end{equation}
where
$x(\tau)$ and $y(\tau)$ are given by Eqs. (\ref{orbit}) and (\ref{io}). 
In a likewise manner as Type 1, we obtain the average input power for Type 2 periodic orbit: 
\begin{equation}
P_{A2} = \frac{q}{T_p}\left(\frac{x_f - X_-}{\alpha} + A_2 \tau_d \right)
\label{pa2}
\end{equation}
These average powers characterize extracting energy from the PV input. 
The theoretical formulae (\ref{pa1}) and (\ref{pa2}) are used in analysis of the BOP. 

Fig. \ref{fg5} shows examples of periodic obits with period $T_p$ and their instantaneous power. 
The contraction rate ($Df_1$, $Df_2$) and dimensionless average input power 
$(P_{A1}, P_{A2})$ are also shown. 
In Type 1, stability of periodic orbit in (a) is weaker 
($Df_1$ is larger ) than that in (b) whereas average  
input power ($P_{A1}$) in (a) is larger than that in (b). 
It suggests existence of a trade-off between stability and average input power. 
The trade-off is analyzed in Section \ref{bop}.

\section{The Biobjective Optimization Problem}
\label{bop}
\begin{figure}[b!]
\centering
\includegraphics[width=0.5\columnwidth]{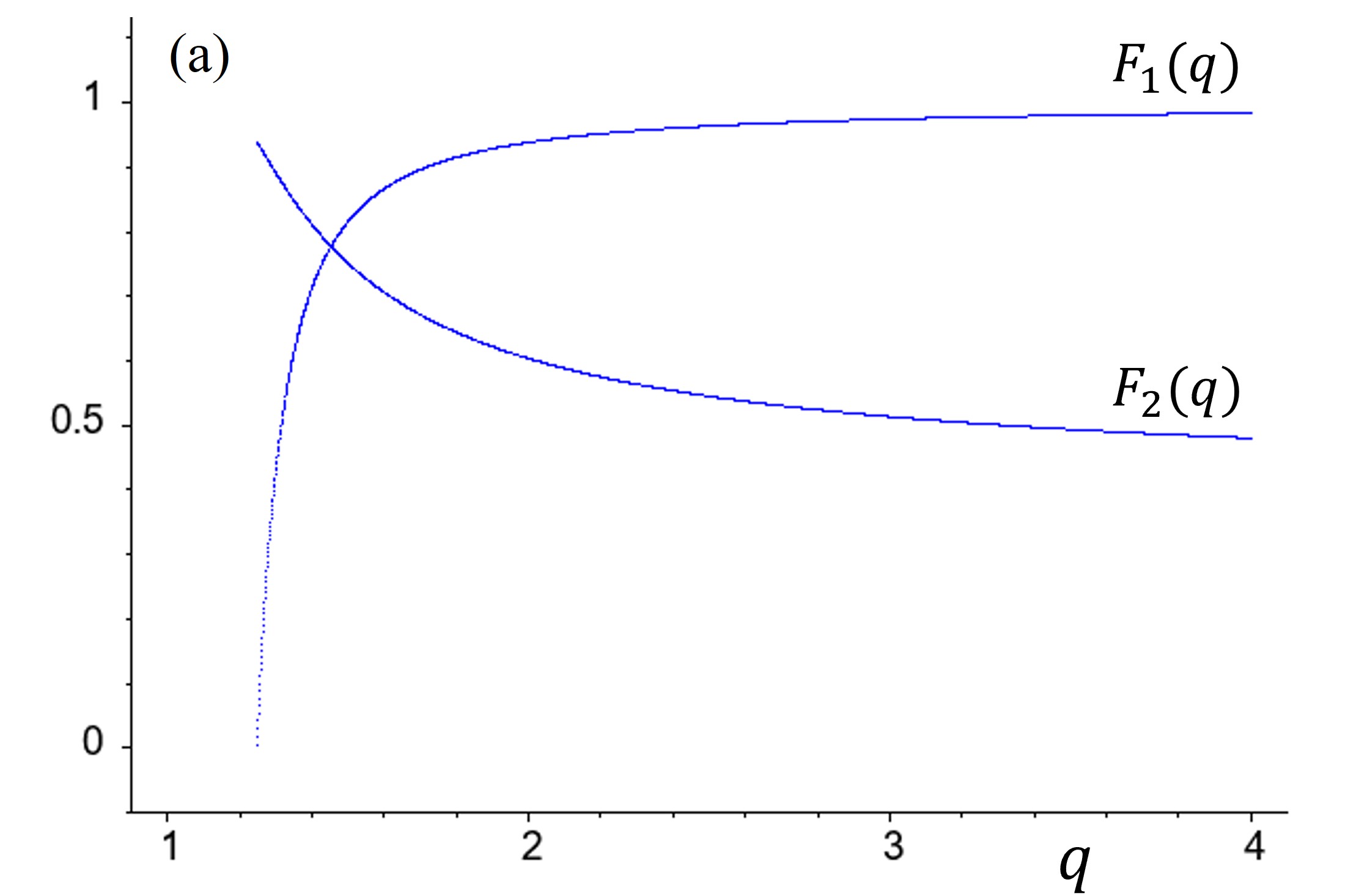}
\includegraphics[width=0.5\columnwidth]{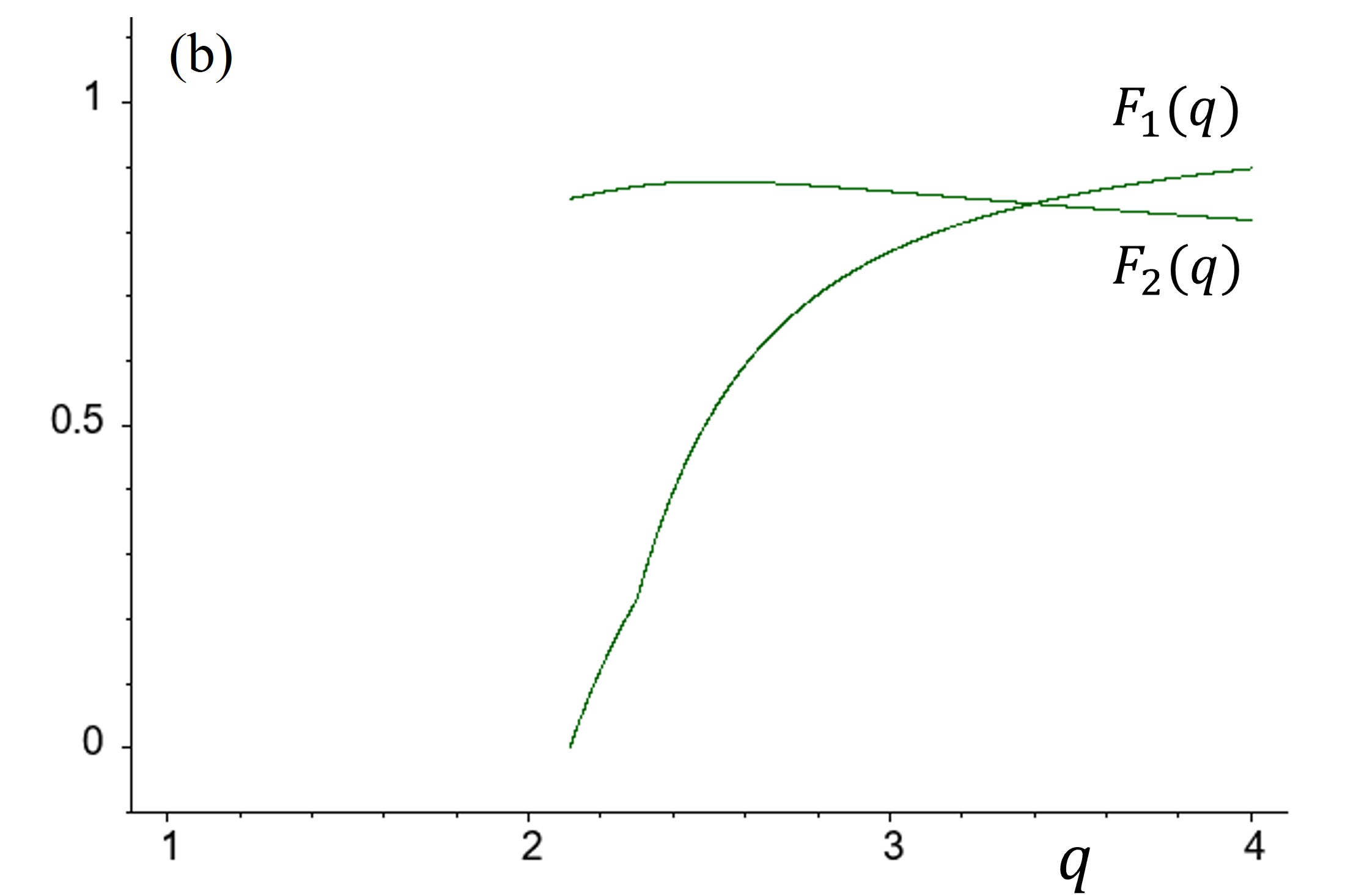}
\includegraphics[width=0.5\columnwidth]{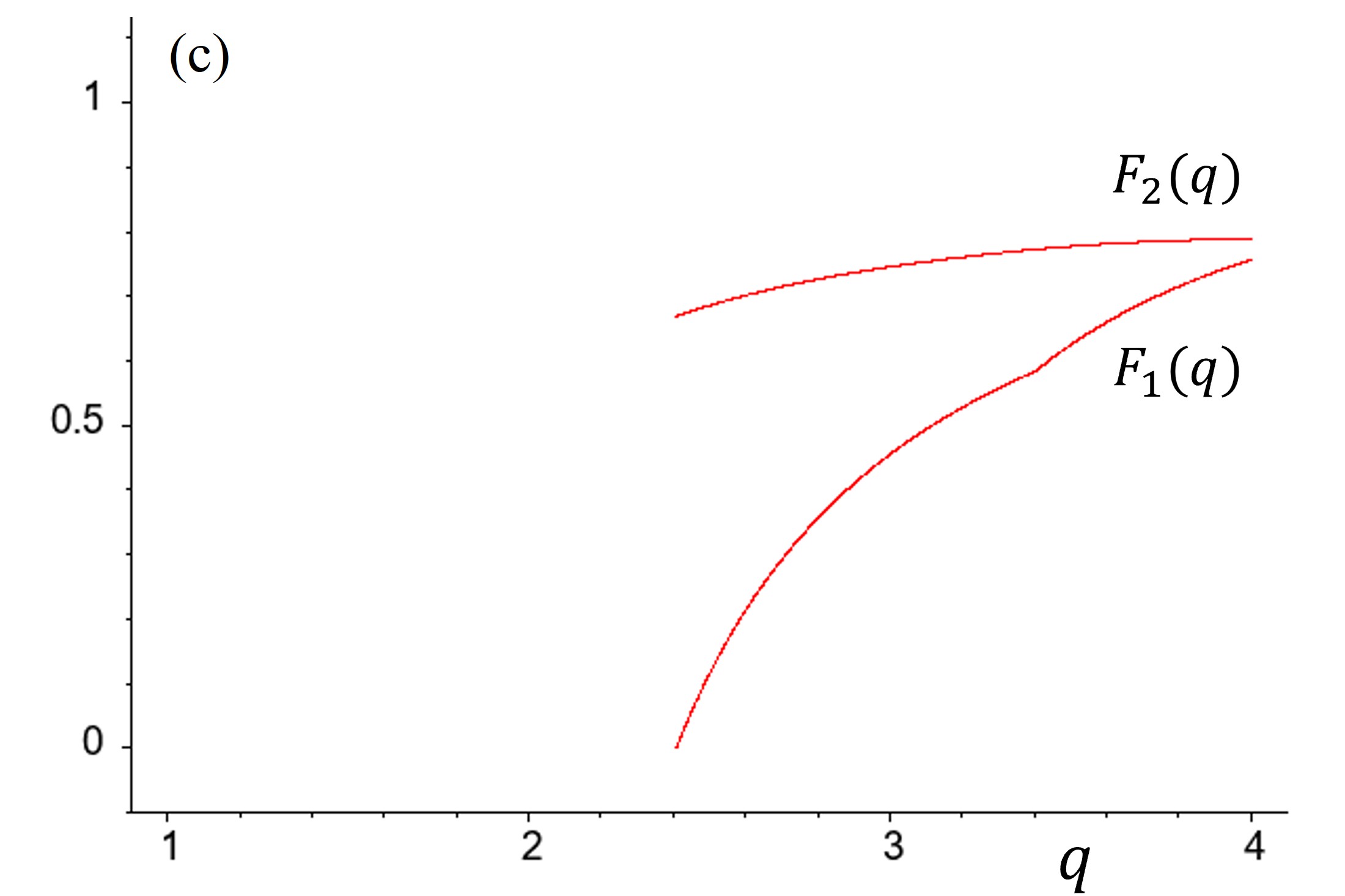}
\caption{Two objectives for $q$. 
(a) $X_-=0.9$, (b) $X_-=0.438$, (c) $X_-=0.2$.}
\label{fg6} 
\end{figure}

This section defines the BOP and calculates the Pareto front exactly. 
Since analysis for all the 5 parameters is not easy (the curse of dimensionality), we select two control parameters $\bm{p}\equiv (q, X_-)$. 
The other three parameters are fixed after trial-and-errors: 
\begin{equation}
\bm{p} \equiv (q, X_-), \ T_p=1, \ \alpha=0.875, \ \beta=3.5.
\label{para}
\end{equation}
We define two objectives as functions of the parameters $\bm{p}$. 
The first objective evaluates stability of the periodic orbits with period $T_p$: 
\begin{equation}
F_1(\bm{p}) = \left\{\begin{array}{ll}
1-|Df_1(\bm{p})| & \mbox{for Type 1}\\
1-|Df_2(\bm{p})| & \mbox{for Type 2}
\end{array}\right.
\label{obj1}
\end{equation}
where $F_1(\bm{p})<1$. 
As $F_1$ increases, stability of the periodic orbit becomes stronger. 
Using the theoretical formulae (\ref{df1}) and (\ref{df2}), $F_1(\bm{p})$ is calculated exactly. 
The second objective evaluates average power of the periodic orbits: 
\begin{equation}
F_2(\bm{p}) = \left\{\begin{array}{ll}
P_{A1}(\bm{p})  & \mbox{for Type 1}\\
P_{A2}(\bm{p})  & \mbox{for Type 2}
\end{array}\right.
\label{obj2}
\end{equation}
where $F_2(\bm{p})<1$. 
As $F_2$ increases, the average input power becomes larger.  
Using the theoretical formulae (\ref{pa1}) and (\ref{pa2}), $F_2(\bm{p})$ is calculated exactly
\footnote{
The theoretical formulae of the two objectives $F_1$ and $F_2$ enable us to calculate the Pareto set exactly.}. 
Fig. \ref{fg6} shows examples of the two objectives for parameter $q$ (dimensionless output voltage) where parameter $X_-$ (dimensionless lower threshold) is fixed. 
We can confirm a trade-off between two objectives in Fig. \ref{fg6} (a) and (b), but cannot confirm it in Fig. \ref{fg6} (c).

\begin{figure}[tb]
\centering
\includegraphics[width=0.5\columnwidth]{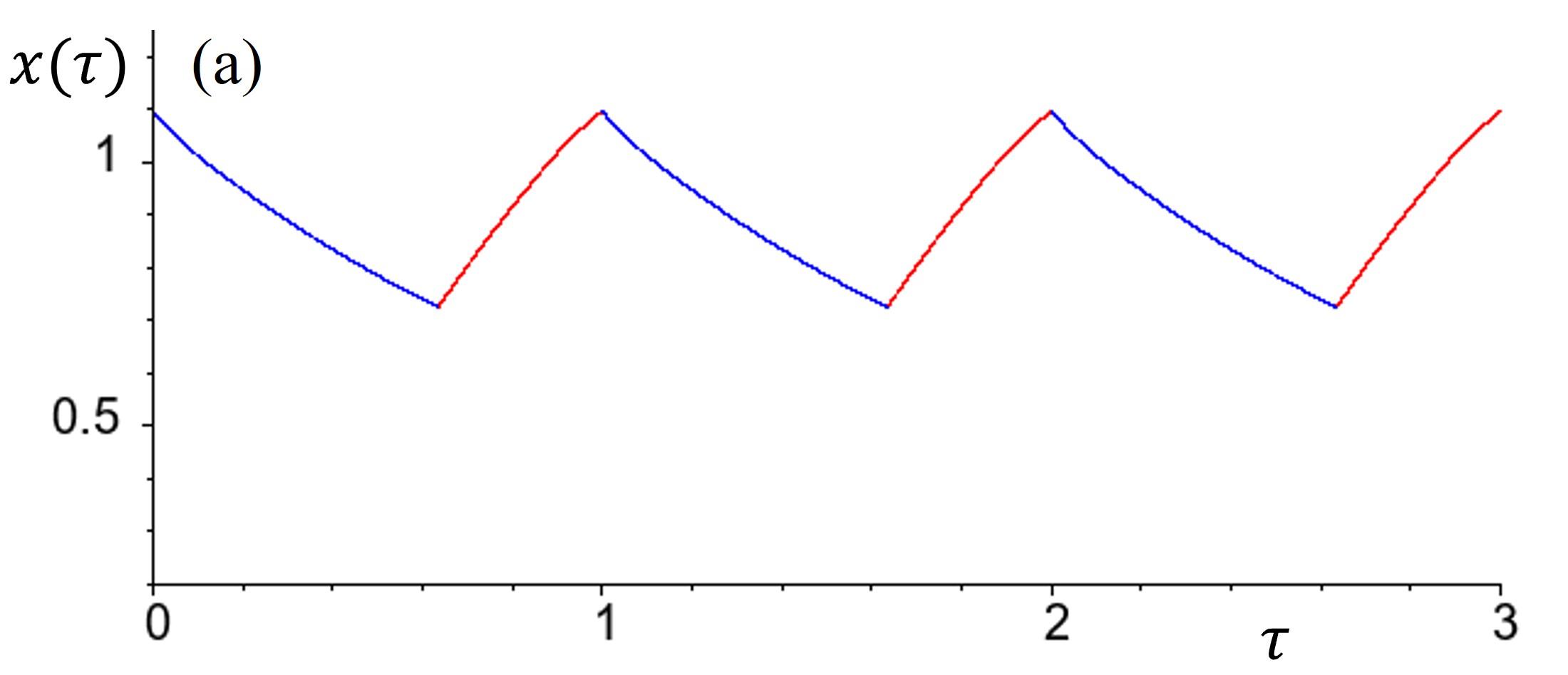}
\includegraphics[width=0.5\columnwidth]{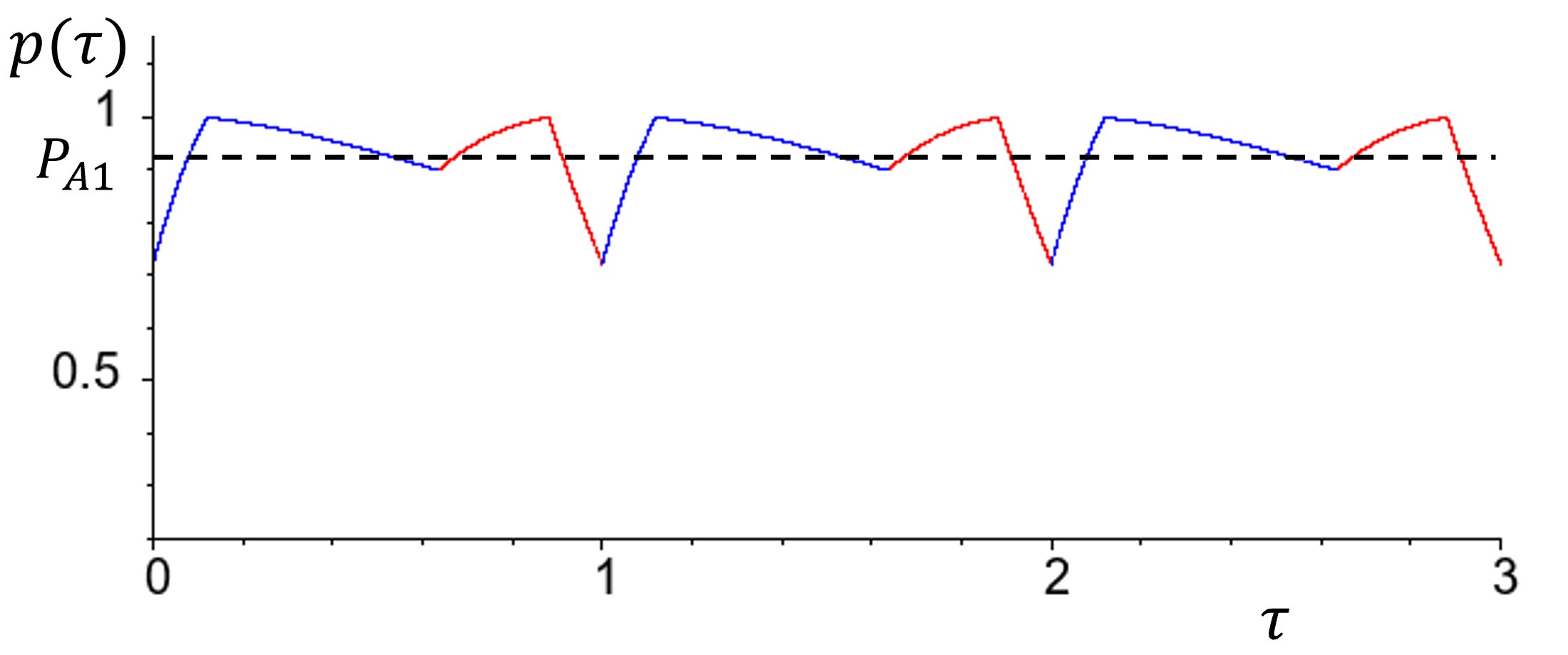}
\includegraphics[width=0.5\columnwidth]{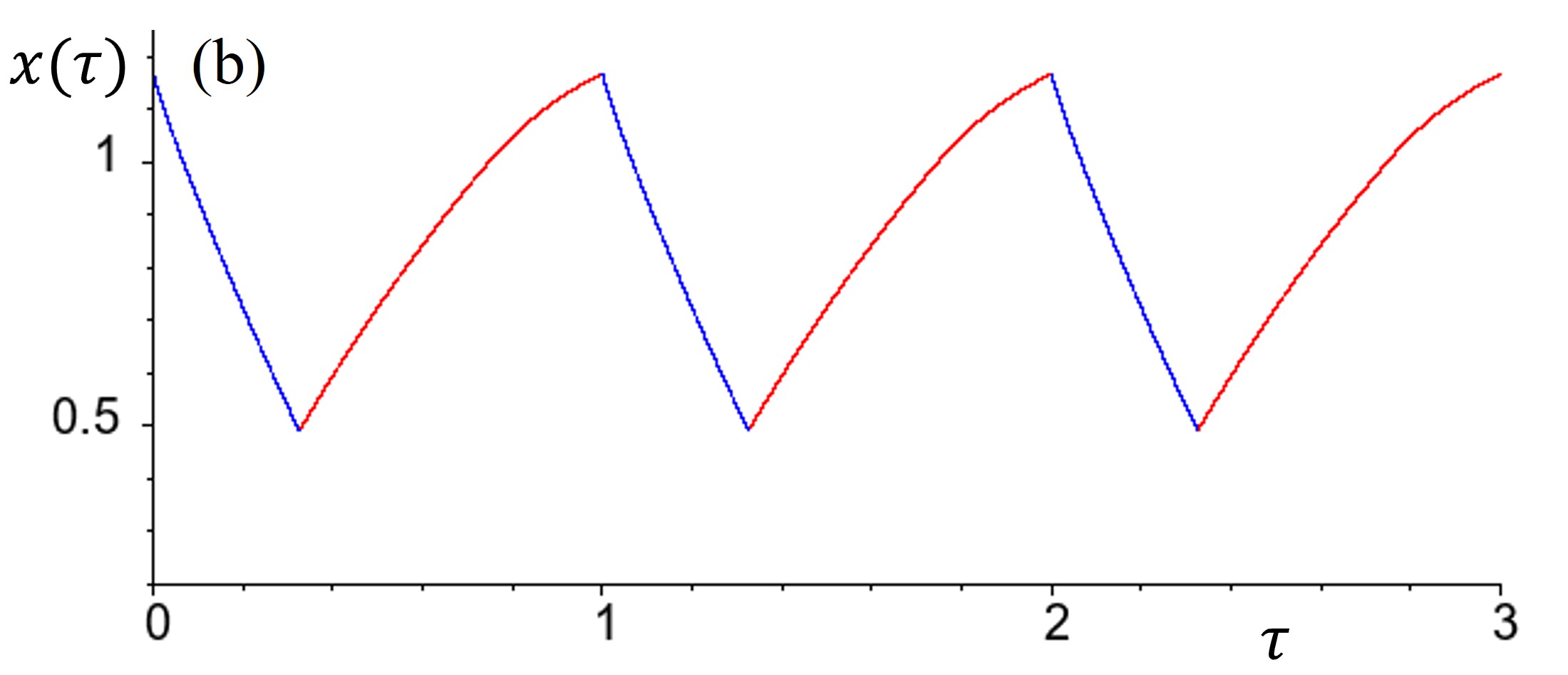}
\includegraphics[width=0.5\columnwidth]{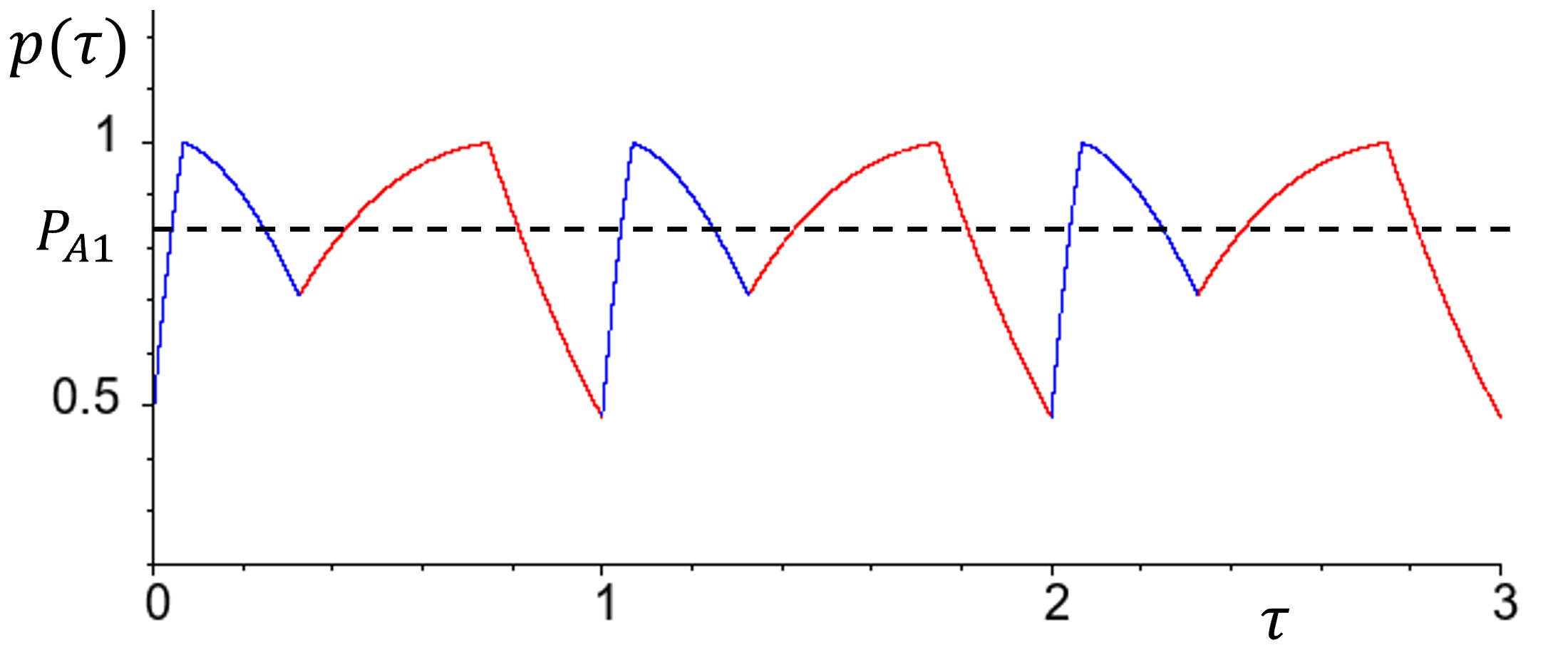}
\includegraphics[width=0.5\columnwidth]{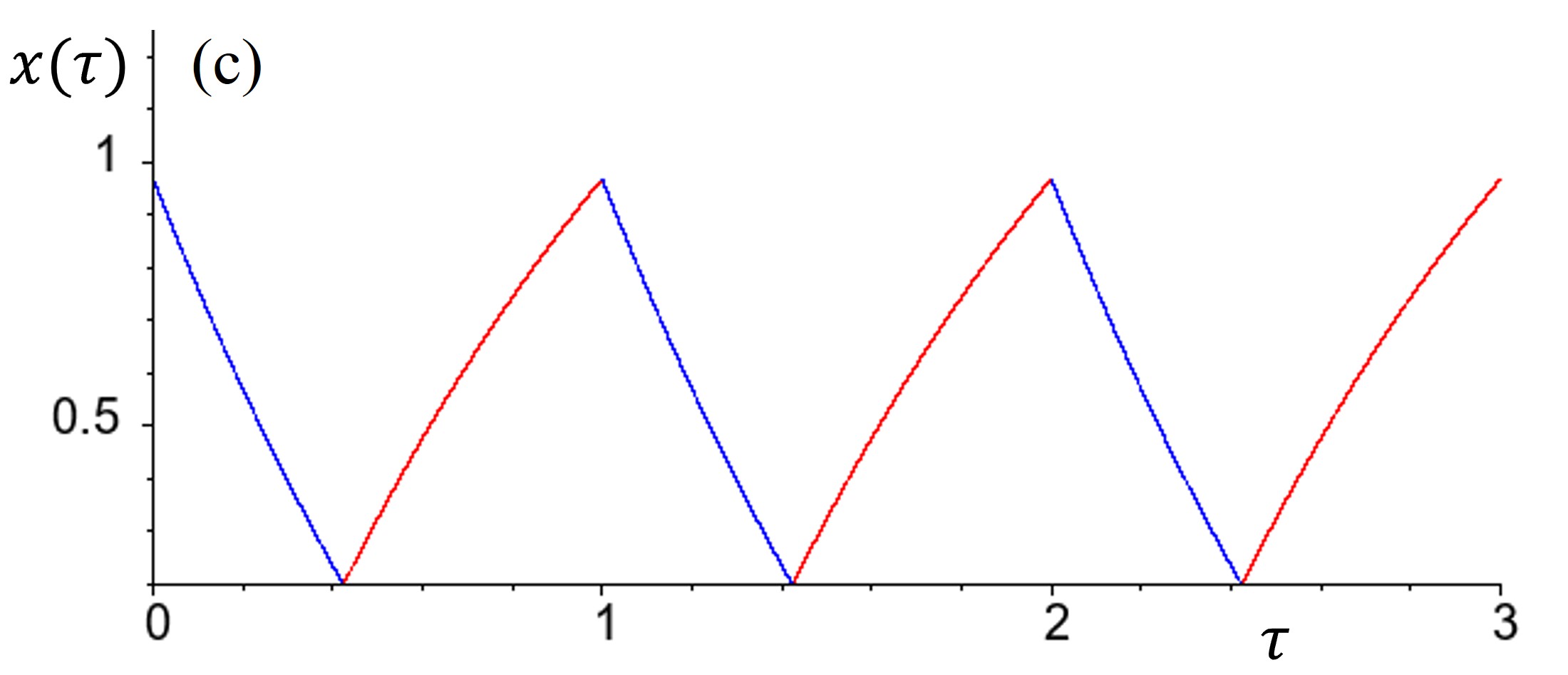}
\includegraphics[width=0.5\columnwidth]{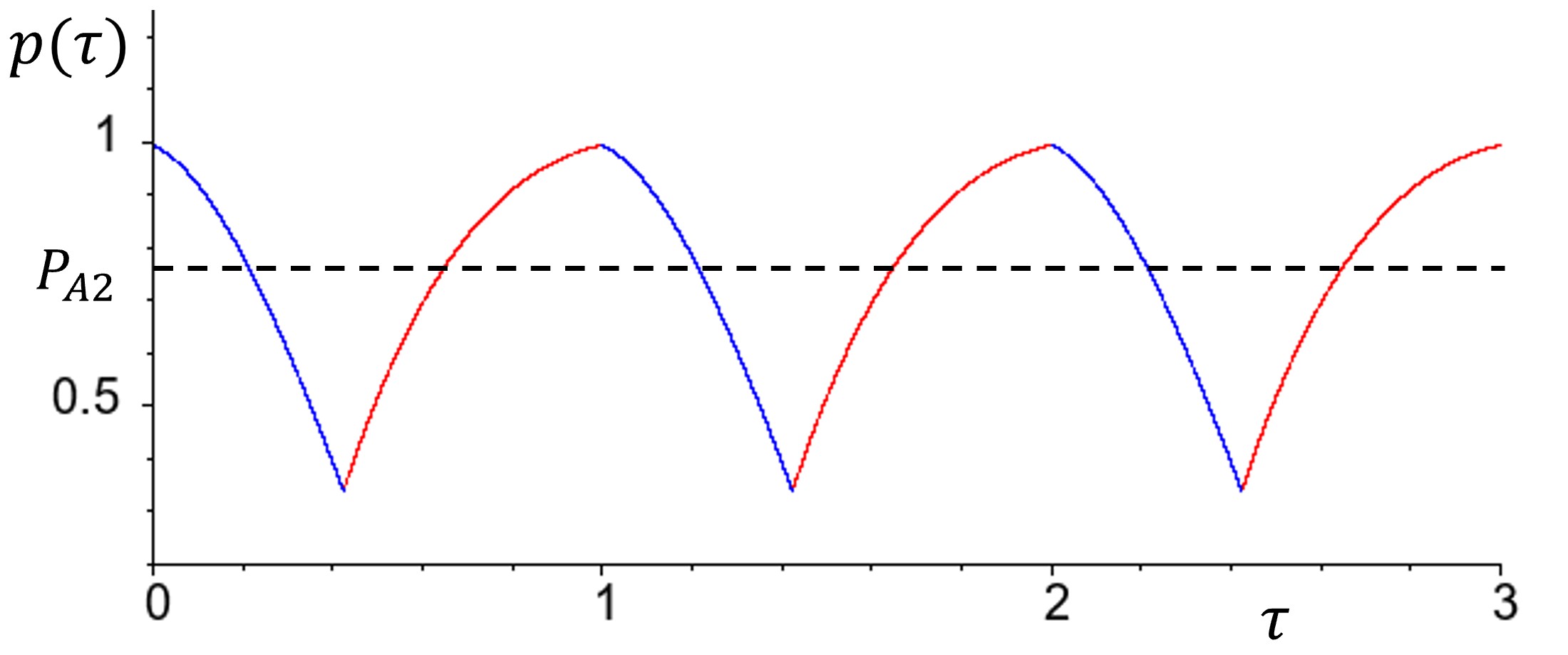}
\caption{Periodic orbit with period $T_p=1$ and instantaneous power for $\alpha=0.875$, $\beta=4\alpha$. 
(a) Type 1 for $X_-=0.726$, $q=1.66$. 
$Df_1=0.658$, $P_{A1}=0.937$ ($F_1=0.342$, $F_2=0.937$). 
(b) Type 1 for $X_-=0.492$, $q=3.2$. 
$Df_1=0.147$, $P_{A1}=0.842$ ($F_1=0.853$, $F_2=0.842$). 
(c) Type 2 for $X_-=0.2$, $q=3.2$. $Df_2=0.528$, $P_{A2}=0.761$
($F_1=0.528$, $F_2=0.761$).}
\label{fg5} 
\end{figure}

\clearpage

\begin{figure}[b!]
\centering
\includegraphics[width=0.7\columnwidth]{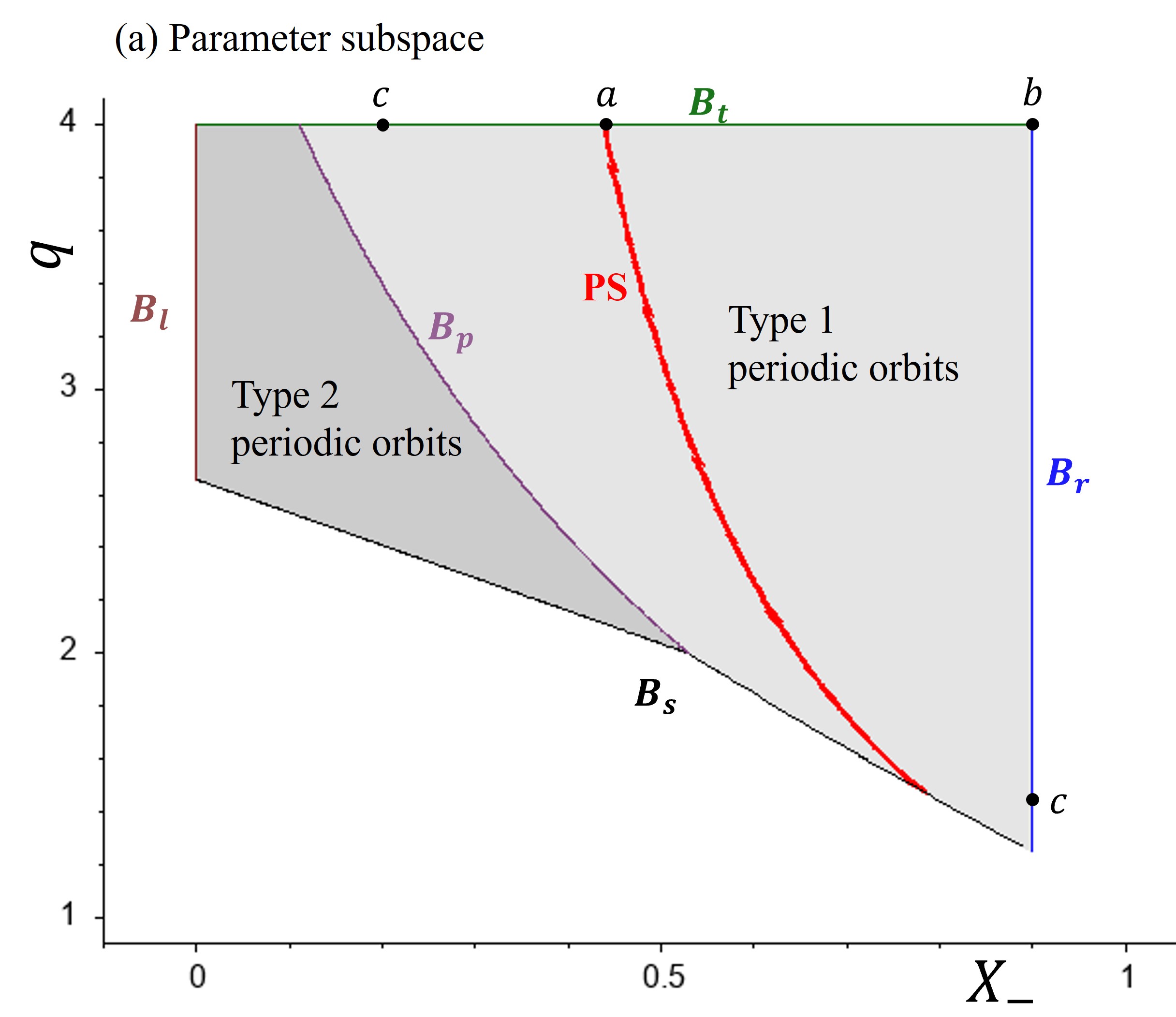}
\includegraphics[width=0.7\columnwidth]{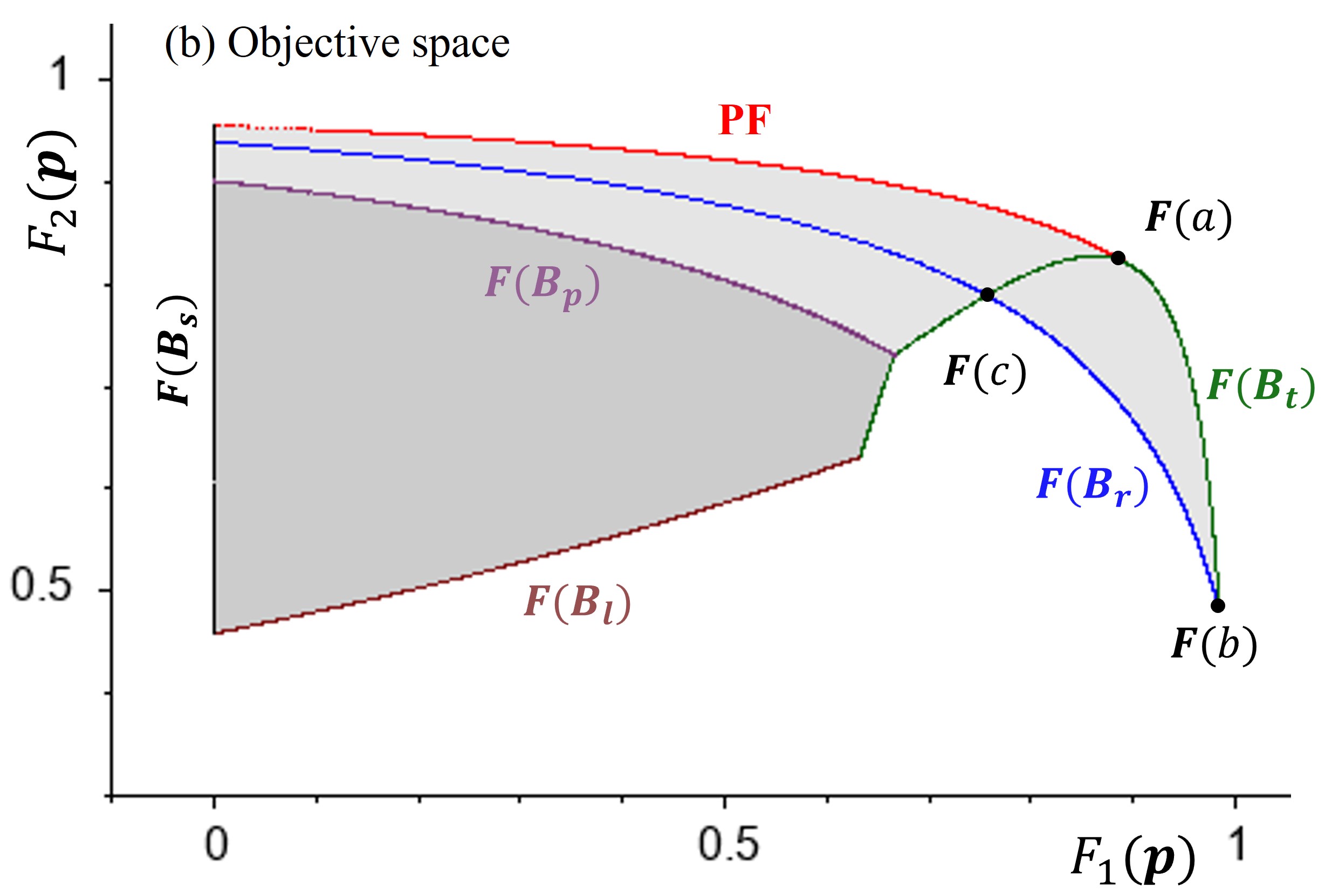}
\caption{Parameter subspace $S_p$ and objective space $S_o$. 
(a) Parameter subspace. $B_s$: border of stable periodic orbit.  
$B_p$: border between Type 1 and Type 2. 
PS: Pareto Set. 
$B_t$, $B_l$, and $B_r$: Top, left, and right edges of $S_p$. 
(b) Objective space. $F(B_s)$: image of $B_s$. 
$F(B_p)$: imaga of $B_p$. 
PF: Pareto front.
$F(B_t)$, $F(B_l)$, and $F(B_r)$: images of $B_t$, $B_l$, and $B_r$. 
}
\label{fg7} 
\end{figure}

\noindent
In order to consider the BOP, we define parameter subspace: 
\begin{equation}
S_p \equiv \{ \bm{p} \ | \ 1 < q < 4, \ 0 < X_- < 0.9, |Df|<1 \}, \ 
Df = \left\{\begin{array}{ll}
Df_1  & \mbox{for Type 1}\\
Df_2  & \mbox{for Type 2}
\end{array}\right.
\label{parasub}
\end{equation}
 In $S_p$, the periodic orbit is stable. 
The range of $q$ and $X_-$ are determined after trial-and-errors. 
Fig. \ref{fg7} (a) shows the parameter subspace with several key borders:
\begin{itemize}

\item $B_p \equiv \{ \bm{p} \ | \ x_f=1 \}$: 
Border between Type 1 and Type 2 periodic orbits. 
The periodic orbit is Type 1 (Type 2) in the right (the left) of this border. 
Using Eq. (\ref{fpoint}), this border is calculated exactly. 

\item $B_s \equiv \{ \bm{p} \ | \ |Df|=1 \}$: 
Border of stable periodic orbit. 
Periodic orbits becomes unstable below this border. 
Using Eqs. (\ref{df1}) and (\ref{df2}), this border is calculated exactly. 

\item $B_t \equiv \{ \bm{p} \ | \ q=4 \}$, 
$B_l \equiv \{ \bm{p} \ | \ X_-=0 \}$, and
$B_r \equiv \{ \bm{p} \ | \ X_-=0.9 \}$: 
Top, left, and right edges of $S_p$. 
These edges and their three points ($a$, $b$, $c$) are used to consider objective space as shown afterward. 
\end{itemize}

Here we define the BOP of the two objectives:
\begin{equation}
    \begin{array}{l}
        \mbox{Maximize $\bm{F}(\bm{p}) \equiv (F_1(\bm{p}), F_2(\bm{p})) \in S_o $}\\
        \mbox{Subject to } \bm{p} \in S_p
    \end{array}
\label{bop1}
\end{equation}
where $S_o$ is the objective space. 
Fig. \ref{fg7} (b) shows the objective space with image of borders in the parameter subspace $S_p$:
$\bm{F}(B_s)$ is the image of $B_s$ and $\bm{F}(B_p)$ is the image of $B_p$. 
($\bm{F}(B_t)$, $\bm{F(}B_l)$, $\bm{F}(B_r)$) are images of ($B_t$, $B_l$, $B_r$) and 
($\bm{F}(a)$, $\bm{F}(b)$, $\bm{F}(c)$) are images of ($a$, $b$, $c$). 
These images are basic to grasp relationship between $S_p$ and $S_o$.  

Applying the brute force attack to investigate the non-dominated relationship in Eq. (\ref{ndom}), we obtain the Pareto set (PS) and Pareto front (PF) exactly as shown in Figs. \ref{fg6} and \ref{fg7}, respectively. 
Note that the non-dominated relationship does not depend on scaling of objectives. 
In the calculation, the theoretical formulae of the objectives (Eqs. (\ref{obj1}), (\ref{df1}), (\ref{df2});  (\ref{obj2}), (\ref{pa1}), (\ref{pa2})) are used for parameter subspace $S_p$ divided into lattice points in size $10^{-3} \times 10^{-3}$. 
Figs. \ref{fg6} and \ref{fg7} are exact main results without approximation. 
We can see the following: 

\begin{itemize}

\item The Pareto front (PF) characterizes the best trade-off and guarantees the existence of the trade-off between two objectives: orbit stability $F_1(\bm{p})$ and average input power $F_2(\bm{p})$
\footnote{
If setting of parameter subspace is not suitable, we cannot obtain the Pareto front.
}. 
In the PF, the periodic orbits are Type 1. Two Type 1 periodic orbits in Fig. \ref{fg5} (a) and (b) are given on the PS/PF: they are an example of a trade-off.  As $S_p$ is divided more finely, e.g. grid size $10^{-4} \times 10^{-4}$, we obtain higher-resolution PF that is an interpolation of the PF in grid size $10^{-3} \times 10^{-3}$.
    
\item The arc from $F(a)$ to $F(b)$ on $F(B_t)$ on $S_o$ is an additional quasi-Pareto front depending on position of the top edge  $B_t$ of the parameter subspace $S_p$. 
The quasi-Pareto front is a part of the Pareto front for $S_p$ with $q \le 4$, however, moves to the right (toward $F_1(\bm{p})=1$) as $q$ increases. 
The PF and quasi-PF are a criterion to consider the trade-off. 

\item In the system design, the strongest stability with the maximum average input power ($F_1=1$ and $F_2=1$) is impossible. 
If a user has priority to efficiency (respectively, to stability), the user selects parameter values for the left (respectively, the right) part of the PF. 

\end{itemize}

\section{Conclusions}
The BOP in the simple piecewise linear SDS has been considered in  this paper. 
The SDS is based on a boost converter with PV input, 
the first objective evaluate stability of periodic orbits, and the second objective evaluate average input power. 
Using the piecewise exact solutions, the two objectives are formulated theoretically. 
Applying the brute force attack to the theoretical formulae on the parameter subspace $S_p$, the Pareto front are obtained exactly on the objective space $S_o$. 
Relationship between $S_p$ and $S_o$ is also considered. 
In our future works, we should consider various problems, including the following. 
\begin{enumerate}
\item 
Applying the brute force attack in 2-dimensional parameter subspace, we have calculated the Pareto set and the Pareto front exactly as shown in Figs. \ref{fg6} and \ref{fg7}. 
However, as the dimension of the parameter space increases, exact calculation becomes impossible (the curse of dimensionality). 
In order to analyze the BOP in higher-dimensional parameter space, we should apply effective multiobjective evolutionary algorithms. 
It should be noted that evolutionary algorithms cannot calculate Pareto front (Pareto set) exactly but give an  approximated Pareto front. 
For example, Ref. \cite{fujikawa} has applied an evolutionary algorithm MOEA/D \cite{moea} to a BOP and has demonstrated an approximated Pareto front. 
Our exact result in Figs. \ref{fg6} \ref{fg7} is a  fundamental step to consider the evolutionary algorithms as suggested in Appendix. 

\item Besides $F_1$ (stability) and $F_2$ (input power), various objectives exist, e.g. response time and attraction basin. MOPs of various objectives are important in the analysis and synthesis of nonlinear dynamical systems. 

\item In MOPs in practical power converters, objective functions have different units/scales.  
In order to explore the Pareto front, efficient scaling/normalization of the objectives is necessary. 

\item In practical PV inputs, the characteristics vary depending on irradiation and temperature \cite{boost1}. 
The MOPs become time-variant and effective analysis method is required. 
\end{enumerate}

\section*{Appendx: }
In order to analyze the BOP in higher-dimensional parameter space, we should apply suitable evolutionary algorithms. 
Here we suggest a fundamental strategy. 
\begin{enumerate}
    \item Applying evolutionary algorithms to the BPO in Eq. (\ref{bop1}), we can obtain approximated Pareto set and Pareto front for 2-dimensional parameter subspace. As popular/classic algorithms, we suggest NSGA-II \cite{nsga}, IBEA \cite{ibea}, and MOEA/D \cite{moea}. 
     In the algorithms, careful hyper-parameter tuning is required \cite{mop1}.  
    \item Comparing the approximated results with exact results in Figs. \ref{fg6} and \ref{fg7}, we evaluate the performance of the evolutionary algorithms. 
    The comparison should be based not only on the approximation error but also on other metrics including the computation cost \cite{hdmo} and the algorithm simplicity. 
     \item Applying suitable evolutionary algorithm(s), we analyze BPOs in higher-dimensional parameter space.  
     The results of the analysis will be developed into an analysis of MOPs in switching power converters with smooth photovoltaic input.   
\end{enumerate}

\end{document}